# First-principles study of phase transition in cadmium titanate by molecular dynamics incorporating nuclear quantum effects

Kansei Kanayama* and Kazuaki Toyoura

*Department of Materials Science and Engineering, Kyoto University, Kyoto 606-8501, Japan*

*kanayama.kansei.67z@st.kyoto-u.ac.jp

**Abstract**

First-principles molecular dynamics (FPMD) simulations were applied for the paraelectric-ferroelectric phase transition in the perovskite-type cadmium titanate, $CdTiO_3$. Since the phase transition is reported to occur at the low temperature around 80 K, the quantum thermal bath (QTB) method was utilized in this study, which incorporates the nuclear quantum effects (NQEs). The structural evolutions in the QTB-FPMD simulations are in reasonable agreement with the experimental results, by contrast in the conventional FPMD simulations using the classical thermal bath (CTB-FPMD). According to our phonon calculations, volume expansion is the key in the stabilization of the ferroelectric phase at low temperatures, which was well reproduced in the QTB-FPMD with the NQEs. Thus, the NQEs are of importance in phase transitions at low temperatures, particularly below the room temperature, and the QTB is useful in that it incorporates the NQEs in MD simulations with low computational costs comparable to the conventional CTB.

# 1. Introduction

The titanium-containing crystals with the perovskite structures, $A$TiO$_3$ ($A$ = Pb, Ba, Cd), are well-known ferroelectric ceramics[1–7]. The paraelectric-ferroelectric phase transitions in these crystals are reported to be the second-order or weak first-order phase transition with little volume gap[6–9]. In the case of the cubic-to-tetragonal transition in PbTiO$_3$, every Ti ion is displaced from the center of the TiO$_6$ octahedral unit to the [001] direction in the pseudo-cubic system, which is originated from the second-order Jahn-Teller effects for the $d^0$ electronic structure of the tetravalent Ti ion[10]. In the case of BaTiO$_3$, three phase transitions (cubic → tetragonal → orthogonal → rhombohedral) occur sequentially with decreasing temperature, where Ti ions are displaced to the [001], [011], and [111] directions, respectively[10].

In theoretical studies, phonon calculations within the harmonic approximation[11] are widely used for elucidating the lattice dynamics of displacive phase transitions[12–18]. The dynamical instabilities of high-temperature phases are evaluated from the vibrational modes, in which imaginary eigenfrequencies correspond to the atomic displacements in displacive phase transitions. Such phonon calculations are basically limited to dynamical stability analysis at 0 K, while first-principles molecular dynamics (FPMD) simulations are able to directly treat phase transition dynamics at finite temperatures, e.g., structural changes and macroscopic dynamical properties around phase transition temperatures. FPMD simulations have therefore been applied for a wide range of perovskite crystals, such as PbTiO$_3$ [19,20], LiNbO$_3$[21], and KNbO$_3$[22].

However, it is difficult to rigorously evaluate lattice dynamics at low temperatures by conventional MD simulations, because the nuclear quantum effects (NQEs), e.g. the zero-point vibration, are not negligible. In the conventional methods, such as the Nose-Hoover thermostat[23,24] or Langevin thermal bath[25], each atom is treated as a classical particle with neglecting the NQEs even in FPMD simulations. To address this issue, Dammak et al. has proposed a new method based on the Langevin MD[26–28], which incorporates the NQEs by using the Quantum Thermal Bath (QTB) instead of the conventional Classical Thermal Bath (CTB). The computational cost of QTB-MD is comparable to that of the CTB-MD, which is the greatest advantage over the huge computational cost of path integral molecular dynamics (PIMD)[29]. The QTB-MD has been applied to the two perovskite-type systems, BaTiO$_3$[30] and SrTiO$_3$[31]. Although these applications are based on not FPMD but the effective Hamiltonian or machine-learning potentials, the QTB-MD simulations showed the reasonable agreement with the PIMD results and reproduced the experimental values more accurately than the CTB-MD.

In this study, we demonstrate the importance of the NQEs in the structural phase transition at low temperatures using the first-principles QTB-MD (QTB-FPMD). We choose perovskite-type CdTiO$_3$ as a model system because it is reported to have the paraelectric-ferroelectric phase transition at $T_c$ ~ 80 K[9,32–36]. The space group of the high-temperature paraelectric phase is identified as *Pnma* (62) by the X-ray and neutron diffraction studies[5,37]. As shown in figure 1, the *Pnma* phase has the structure with a $a^+b^-b^-$ distortion in the Glaser notation, in which the TiO$_6$ octahedra are rotated alternately around the two of the three [001] axes in the pseudo-cubic phase, and simultaneously around the other axis. In contrast, the crystal structure of the low-temperature ferroelectric phase is still in controversy. The two crystal structures with the space groups *Pna*2$_1$ (33)[32,37] and *P*2$_1$*ma* (26)[38] have been suggested as the ferroelectric phase in the X-ray and neutron diffraction studies. The space groups of both phases are the non-centrosymmetric subgroups of the space group *Pnma*, and the difference is a symmetry element, a diagonal gride plane or a mirror plane. In the previous first-principles studies, the phonon calculations were performed for the paraelectric and ferroelectric phase[16,17], concluding that the ferroelectric phase is likely to have the space group *Pna*2$_1$ rather than *P*2$_1$*ma* by comparing the magnitudes of the eigenfrequencies between imaginary vibrational modes. FPMD simulations for CdTiO$_3$ have not been reported, probably because the CTB-FPMD is not suitable for analyzing phase transition dynamics around the low phase transition temperature. In this study, we have investigated the temperature dependences in the cell volume and lattice constants by the QTB-FPMD simulations, analyzing the structural evolutions around the transition temperature. We also applied the CTB-FPMD for comparison, demonstrating the importance of incorporating the NQEs in the MD simulations. In addition, the phonon calculation for each phase was performed for elucidating the origin of the superiority in the QTB-FPMD over the CTB-FPMD. The methodology of the QTB method and the computational conditions are presented in Section II, and the results and discussions of the applications are presented in Section III.

## 2. Computational method

### 2.1. Quantum thermal bath in Langevin MD

The QTB-MD method is a modification of the Langevin MD[25] for incorporating the NQEs. It is based on the Langevin equation expressed as

$$m_i(d^2 x_{i\alpha}/dt^2) = - dU/dx_{i\alpha} - m_i\gamma(dx_{i\alpha}/dt) + R_{i\alpha}(t), \qquad (1)$$

where $m_i$ and $x_{i\alpha}$ are the mass and three-dimensional cartesian coordinate ($\alpha$ = 1, 2, 3) of the $i$th atom, respectively, and $t$, $U$, and $\gamma$ are the time, the potential energy, and the friction coefficient, respectively. The last two terms of the right-hand side in Eq. (1) correspond to the friction and random forces, respectively, which are fictitious forces controlling the temperature. In the CTB-based Langevin MD, CTB-MD, a time-uncorrelated random force is given as a Gaussian white noise for enabling the system to evolve under the energy equipartition theorem. This theorem means that the mean kinetic energy is 1/2 $k_B T$ per degree of freedom, where $k_B$ and $T$ are the Boltzmann constant and the temperature, respectively.

In the QTB-MD[26–28], the kinetic energy is controlled for including the NQEs, while every atom in a given system is treated as a classical particle. A time-correlated random force is used instead of the time-uncorrelated force in the CTB-MD. The power spectral density $I_R(\omega)$ of the time-correlated random force is expressed as

$$I_R(|\omega|) = 2m\gamma \, [1/2 \, \hbar\omega + \hbar\omega/(\exp(\hbar\omega/k_B T) - 1)], \qquad (2)$$

where $\hbar$ and $\omega$ are the Dirac constant and the angular frequency, respectively. Under the harmonic approximation, the mean energy of the system, $E$, corresponds to the quantum harmonic energy of the lattice vibration,

$$E = \Sigma_i \, [1/2 \, \hbar\omega_i + \hbar\omega_i/(\exp(\hbar\omega_i/k_B T) - 1)], \qquad (3)$$

where $\omega_i$ is the eigenfrequency of the $i$th vibrational mode and the first term corresponds to the zero-point energy.

### 2.2. First principles calculations

The DFT-based calculations implemented in the VASP package[39–41] were employed in the structural optimizations, phonon calculations, and MD simulations for the model system of $CdTiO_3$. The projector-augmented wave (PAW) method[42] was used with the cut off energy of the plane-wave as 500 eV. The pseudopotentials were used, in which the following electrons are explicitly treated as valence electrons; 4d and 5s for Cd, 4s and 3d for Ti, and 2s and 2p for O. The local density approximation (LDA)[43,44] was used as the exchange-correlation functional because the LDA is reported to reproduce the experimentally-determined lattice constants of $CdTiO_3$ more accurately[17] than the generalized gradient approximation with Perdew-Burke-Ernzerhof parametrization (PBE_GGA)[45]. The obtained lattice constants under the LDA and the PBE_GGA are shown in table S1 of Supplemental Materials, which coincide with those in the previous DFT study[17] within the computational accuracy.

The structural optimizations were performed for the three polymorphs, the paraelectric phase with the space group *Pnma* and the two ferroelectric phases with *Pna*$2_1$ and *P*$2_1$*ma*. The *k*-point mesh for Brillouin zone sampling in each unit cell was set to 4×2×4 for *Pnma*, 4×4×2 for *Pna*$2_1$, and 2×4×4 for *P*$2_1$*ma*. The structural optimizations were terminated when the norms of all the forces are smaller than 1×10$^{-4}$ eV/Å. After the structural optimizations, the phonon calculations were performed by the finite displacement method implemented in phonopy[46,47]. The supercells consisting of 2×1×2, 2×2×1, and 1×2×2 unit cells were used for the three polymorphs with *Pnma*, *Pna*$2_1$, and *P*$2_1$*ma*, respectively. The *k*-points mesh for these supercells was set to 2×2×2.

Since the CTB-based Langevin FPMD simulations are already implemented in the VASP package, we added the QTB on the VASP code. Specifically, we additionally implemented the code for calculating the time-correlated random force at each MD step. The *NPT*-MD simulations were performed on the CdTiO$_3$ systems with the QTB or the already-implemented CTB for comparison. The temperature was set to 25, 50, 75, 100, 125, 150, 200, 250, or 300 K at each simulation, and the pressure was controlled to the ambient condition using the Parinello-Rahman barostat[48]. The time step and the total simulation time were 1 fs and 100 ps, respectively, starting from the 2×1×2 supercell of the *Pnma* structure. As for the Langevin thermal bath parameters, the friction coefficient $\gamma$ in Eq. (1) was set to 5 THz. The domain of the frequency $\omega$ in Eq. (2) was set to [−314, 314] in THz to avoid the power spectral density $I_R(\omega)$ diverging. The kinetic energies, volumes, and lattice constants in each simulation were averaged during 95 ps after 5 ps for thermal equilibration.

## 3. Results and Discussion

First of all, the structural optimizations were performed for the three CdTiO$_3$ polymorphs with the space group of *Pnma*, *Pna*2$_1$, and *P*2$_1$*ma*. As a result, all the three structures converged to the paraelectric *Pnma* structure, indicating that the high temperature paraelectric phase is the ground state under our computational condition. The optimized volume is 213.82 Å$^3$, and the lattice constant *a*, *b*, and *c* are 5.38, 7.56, and 5.25 Å, respectively. Figure 2(a) shows the calculated phonon band structure of the *Pnma* structure. No imaginary mode means that the paraelectric structure is dynamically stable. The obtained phonon band structure is consistent with the one in the previous DFT study under the metaGGA[18]. However, it is inconsistent with the one with imaginary modes at the Γ-point in the other previous study under the LDA[16,17]. The discrepancy is probably due to the different pseudopotentials and computational conditions.

Note that the dynamical stability of the paraelectric structure is sensitive to the cell volume. Figure 2(b) shows the calculated phonon band structure of the paraelectric *Pnma* phase with expanding each lattice constant by 0.3 %, corresponding to the 0.9 % expansion in the cell volume. The slight volume expansion softens the two vibrational modes at the Γ-point, which have imaginary eigenfrequencies, 0.94*i* and 1.29*i* THz, respectively. The two ferroelectric phases with *Pna*2$_1$ and *P*2$_1$*ma* were obtained by optimizing the displaced structure along the direction of each imaginary vibration mode. This dynamical instability due to the volume expansion of the paraelectric phase was also reported in the previous studies under the metaGGA[18] and the LDA[17]. In our phonon calculations of the paraelectric phase within PBE_GGA, there are four imaginary vibrational modes at the Γ-point in the phonon band structure shown in figure S1, probably due to the relatively large cell volume to that under the LDA.

QTB- and CTB-FPMD simulations were then performed using the *Pnma* structure as the initial structure. Note the computational cost of each MD simulations at each temperature is shown in Table S2 in Supplemental Materials, indicating the computational costs in the QTB-FPMD are comparable to those in the CTB-FPMD. Figure 3(a) shows the mean kinetic energies during the QTB- or CTB-FPMD simulations as a function of temperature. The quantum and classical kinetic energies under the harmonic approximation of the lattice vibrations are also shown in figure 3(a), which were estimated from the eigenfrequencies of the structure *Pnma* obtained by the finite displacement method and from the energy equipartition theorem, respectively. The mean kinetic energies during the QTB- and CTB-FPMD simulations are in reasonable agreement with the quantum and classical ones under the harmonic approximation, respectively. The kinetic energies are larger in the QTB-FPMD than those in the CTB-FPMD, particularly at low temperatures, mainly due to the zero-point vibrations. Therefore, the NQEs are not negligible in the phase transition of CdTiO$_3$.

Figure 4 shows the nuclear density distributions of oxide ions in the *ab*-plane during the QTB- and CTB-FPMD simulations at the lowest temperature, 25 K (See figure S3-S5 in Supplemental

Materials for detailed nuclear density distributions of Cd, Ti, and O ions). In the CTB-FPMD, the distributions of O ions at the two sites of the structure *Pnma* are strongly localized, while the O ions are widely distributed in the QTB-FPMD. The difference in the nuclear density distributions is originated from the high kinetic energy even at the low temperature in the QTB-FPMD, which corresponds to the zero-point vibrations.

The zero-point vibrations also have effects on the mean cell volume in the QTB-FPMD, as shown in figure 3(b). In the CTB-FPMD, the mean cell volume almost linearly depends on the temperature, while the cell volume non-linearly increases with temperature in the QTB-FPMD, nearly constant in the low-temperature range below 100 K. The cell volume in the QTB-FPMD is relatively large to that in the CTB-FPMD. For example, the cell volume at 25 K in the QTB-FPMD was about 0.8 % larger than that of the optimized structure with *Pnma*, while only 0.07 % larger in the CTB-FPMD. The relatively large cell volume can induce the instability of the paraelectric structure with *Pnma*, leading to the stabilization of the ferroelectric phase at low temperatures in the QTB-FPMD. In contrast, the high-temperature paraelectric phase remains stable even at low temperatures in the CTB-FPMD because of the almost same cell volume as the optimized structure.

Figure 5 shows the comparison of the cell volume and lattice constants between our QTB-FPMD, our CTB-FPMD, and the previous experimental study[18]. All data is normalized by the value at the lowest temperature. Regarding the cell volume, there are large discrepancies between the CTB-FPMD and the experiment. The cell volume in the CTB-FPMD linearly increases with increasing temperature, and the change in slope cannot be observed clearly. Considering that the cell volume at 25 K in the CTB-FPMD is almost the same as the one in the optimized *Pnma* structure, and that our phonon calculations indicate the *Pnma* structure is dynamically stable at the optimized volume, the paraelectric phase should be always stable in this temperature range. By contrast, the cell volume in the QTB-FPMD shows comparable temperature dependence to that in the experiment, which have a clear change in slope around the reported phase transition temperature[9,32–36].

As in the cell volumes, the QTB-FPMD reproduced the temperature dependences of the experimental lattice constants more accurately than the CTB-FPMD. In the experiments[9,18], the negative thermal expansion of the lattice constant *a* is reported around the phase transition temperature. This characteristic temperature dependence is qualitatively reproduced in the QTB-FPMD. The lattice constant decreases under 50 K with increasing temperature in the QTB-FPMD as in the experiments. In contrast, the CTB-FPMD shows linear thermal expansion under 100 K, probably due to the stable paraelectric phase even at the lowest temperature. Although the large fluctuations of the lattice constants in these simulations make it difficult to obtain the fully precise temperature dependences, the reasonable results in the QTB-FPMD indicates the importance of the NQEs in the structural phase transition at low temperatures.

## 4. Conclusion

The QTB-FPMD were performed for the paraelectric-ferroelectric phase transition in the perovskite-type $CdTiO_3$. In the QTB method, the NQEs are incorporated by using a random force with time correlations instead of a time-uncorrelated Gaussian random force in the CTB-based Langevin MD. The method was additionally coded on VASP, where the CTB-FPMD are already implemented.

The QTB-FPMD exhibited relatively large kinetic energies to those in the CTB-FPMD when compared at the same temperature. This is because the QTB-FPMD incorporates the NQEs through the kinetic energies. The NQEs are mainly due to the zero-point vibration, which lead to larger cell volumes than those without the NQEs, particularly at low temperatures. In comparison with the previous experimental results, the QTB-FPMD were in reasonable agreement with the reported temperature dependences of the cell volume and lattice constants. In the QTB-FPMD, non-linear thermal expansion can be seen at low temperatures as in the experiments. By contrast, the CTB-FPMD showed linear thermal expansion also in the low-temperature region ($T < 100$ K). The phonon calculations indicate that the paraelectric phase was dynamically stable at 0 K, meaning no paraelectric-ferroelectric phase transition in the CTB-FPMD within the LDA. The ferroelectricity at low temperatures in the QTB-FPMD was originated from larger cell volume than that in the CTB-FPMD, which lead to the dynamical instability in the paraelectric phase. Considering that the QTB-FPMD reproduced the experimental structural parameters more accurately than the CTB-FPMD, the NQEs are of importance for the phase transition in the $CdTiO_3$ system.

**Acknowledgements**

This paper was partially supported by JSPS, KAKENHI (Grant No. 19H05787), and JST, the establishment of university fellowships towards the creation of science technology innovation (Grant No. JPMJFS2123). The supercomputer of ACCMS, Kyoto University was used for FPMD simulations. K.K. was financially supported by Fujinomori-Masamichi Memorial Scholarship in the Mining and Materials Processing Institute of Japan.

[19] Srinivasan V, Gebauer R, Resta R and Car R 2003 *AIP Conf. Proc.* **677** 168

[20] Fang H, Wang Y, Shang S and Liu Z-K 2015 *Phys. Rev. B* **91** 024104

[21] Sanna S and Schmidt W G 2012 *IEEE Trans. Ultrason., Ferroelect., Freq. Contr.* **59** 1925

[22] Tan Z, Peng Y, An J, Zhang Q and Zhu J 2021 *Inorg. Chem.* **60** 7961

[23] Nosé S 1984 *The Journal of Chemical Physics* **81** 511

[24] Nosé S 1991 *Progress of Theoretical Physics Supplement* **103** 1

[25] Grest G S and Kremer K 1986 *Phys. Rev. A* **33** 3628

[26] Dammak H, Chalopin Y, Laroche M, Hayoun M and Greffet J-J 2009 *Phys. Rev. Lett.* **103** 190601

[27] Dammak H, Hayoun M, Brieuc F and Geneste G 2018 *J. Phys.: Conf. Ser.* **1136** 012014

[28] Barrat J-L and Rodney D 2011 *J Stat Phys* **144** 679

[29] Marx D and Parrinello M 1996 *J. Chem. Phys.* **104** 4077

[30] Brieuc F, Bronstein Y, Dammak H, Depondt P, Finocchi F and Hayoun M 2016 *J. Chem. Theory Comput.* **12** 5688

[31] Wu H, He R, Lu Y and Zhong Z 2022 *Phys. Rev. B* **106** 224102

[32] Sun P-H, Nakamura T, Shan Y J, Inaguma Y and Itoh M 1998 *Ferroelectrics* **217** 137–45

[33] Shan Y J, Mori H, Imoto H and Itoh M 2002 *Ferroelectrics* **270** 381

[34] Guzhva M E, Lemanov V V and Markovin P A 2001 *Phys. Solid State* **43** 2146

[35] Torgashev V I, Yuzyuk Y I, Shirokov V B, Lemanov V V and Spektor I E 2005 *Phys. Solid State* **47** 337

[36] Gorshunov B P, Pronin A V, Kutskov I, Volkov A A, Lemanov V V and Torgashev V I 2005 *Phys. Solid State* **47** 547–55

[37] Kennedy B J, Zhou Q and Avdeev M 2011 *Journal of Solid State Chemistry* **184** 2987–93

[38] Shan Y J, Mori H, Tezuka K, Imoto H and Itoh M 2003 *Ferroelectrics* **284** 107–12
10

**Table and Figure captions**

Figure 1. (a) Crystal structure of CdTiO$_3$ paraelectric phase with the space group of *Pnma*. (b) $a^+$ and (c) $b^-$ distortions in the pseudo-cubic systems.

Figure 2. Calculated phonon band structures of the *Pnma* structures (a) with the optimized cell volume and (b) with 0.3 % expanded cell volume.

Figure 3. (a) Kinetic energies and (b) cell volumes during the QTB-FPMD (red circle) and the CTB-FPMD (blue square) as a function of temperature. The quantum and classical kinetic energies under the harmonic approximation of the lattice vibrations are also shown in the red solid and blue dashed lines, respectively. Error bars correspond to a standard deviation in each simulation.

Figure 4. Nuclear density distributions of (a, b) O1 and (c, d) O2 sites in the *ab*-plane during the QTB- and CTB-FPMD simulations at 25 K. The distributions were obtained by accumulating Gaussian distributions centered at each O ion with width of 0.05 Å at every MD step.

Figure 5. The normalized (a) cell volumes and (b-d) lattice constants as a function of temperature. The data in the QTB-FPMD (red circle), CTB-FPMD (blue square), and experiments (black cross) are normalized at the lowest temperature. The experimental data are obtained by scanning the figure in [18] and the phase transition temperature reported in the previous experiments [9,32–36] are shown in a green line.

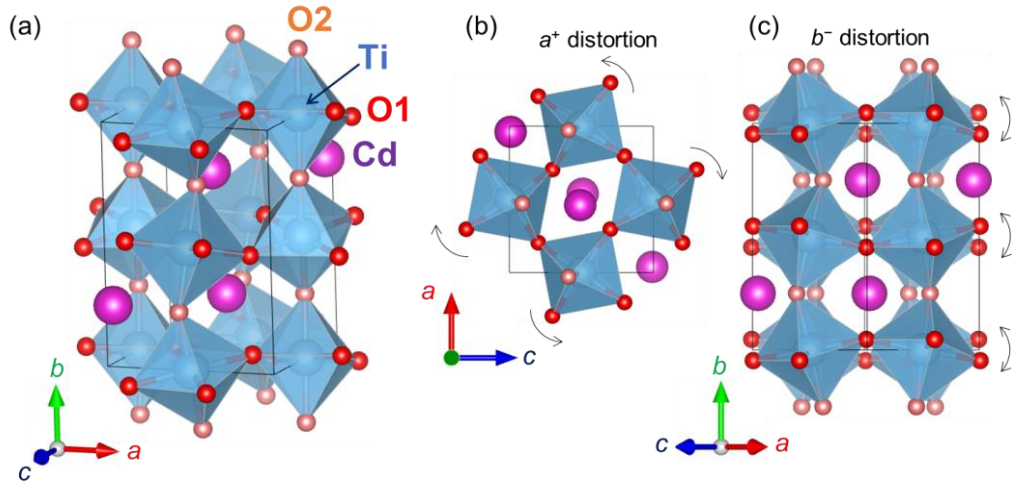

Figure 1. (a) Crystal structure of CdTiO$_3$ paraelectric phase with the space group of *Pnma*. (b) $a^+$ and (c) $b^-$ distortions in the pseudo-cubic systems.

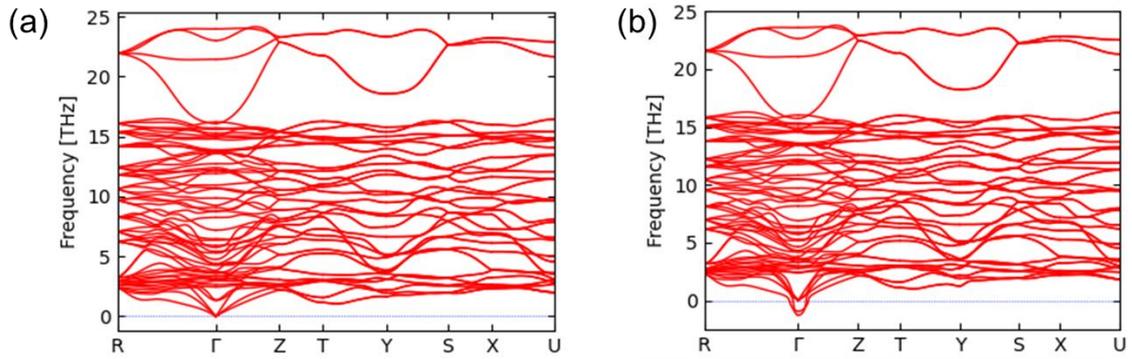

Figure 2. Calculated phonon band structures of the *Pnma* structures (a) with the optimized cell volume and (b) with 0.3 % expanded cell volume.

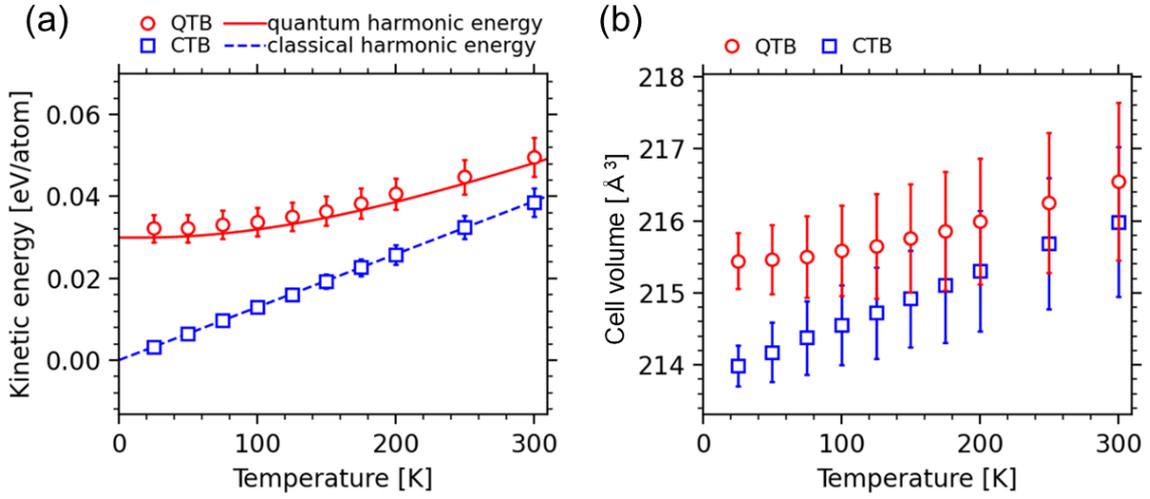

Figure 3. (a) Kinetic energies and (b) cell volumes during the QTB-FPMD (red circle) and the CTB-FPMD (blue square) as a function of temperature. The quantum and classical kinetic energies under the harmonic approximation of the lattice vibrations are also shown in the red solid and blue dashed lines, respectively. Error bars correspond to a standard deviation in each simulation.

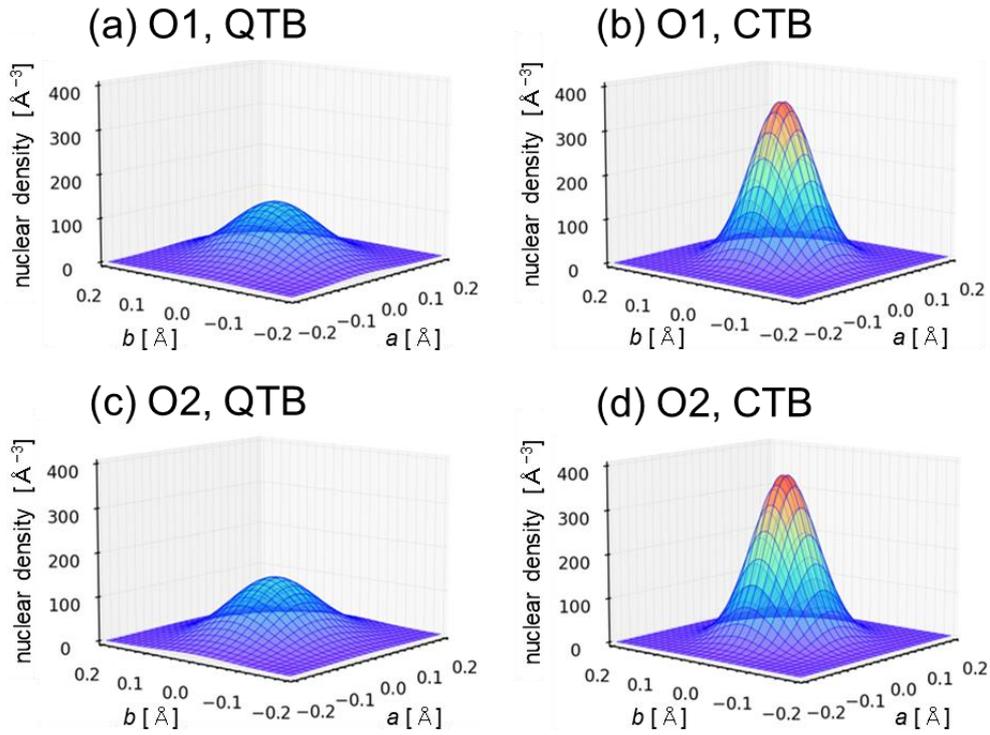

Figure 4. Nuclear density distributions of (a, b) O1 and (c, d) O2 sites in the *ab*-plane during the QTB- and CTB-FPMD simulations at 25 K. The distributions were obtained by accumulating Gaussian distributions centered at each O ion with width of 0.05 Å at every MD step.

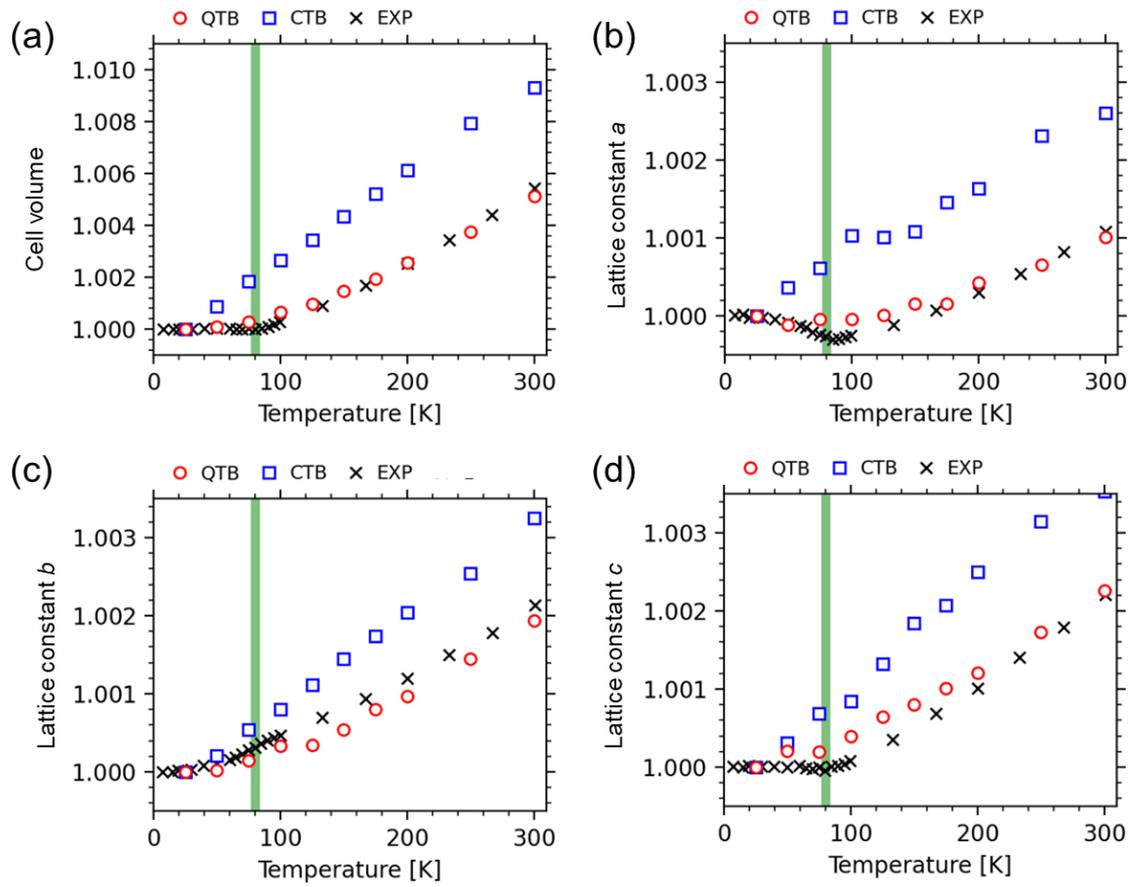

Figure 5. The normalized (a) cell volumes and (b-d) lattice constants as a function of temperature. The data in the QTB-FPMD (red circle), CTB-FPMD (blue square), and experiments (black cross) are normalized at the lowest temperature. The experimental data are obtained by scanning the figure in [18] and the phase transition temperature reported in the previous experiments [9,32–36] are shown in a green line.

# First-principles study of phase transition in cadmium titanate by molecular dynamics incorporating nuclear quantum effects

Kansei Kanayama* and Kazuaki Toyoura

*Department of Materials Science and Engineering, Kyoto University, Kyoto 606-8501, Japan*

*kanayama.kansei.67z@st.kyoto-u.ac.jp

Table S1. Lattice constants *a*, *b*, and *c* of the structure *Pnma* optimized by LDA and PBE_GGA.

|  | *a* [Å] | *b* [Å] | *c* [Å] |
|---|---|---|---|
| **LDA** | 5.38 | 7.56 | 5.25 |
| **PBE_GGA** | 5.49 | 7.72 | 5.37 |

Table S2. The average cpu times (in second / step) in the QTB- and CTB-FPMD simulations during 10000-20000 steps. All MD calculations were performed on the machine with Intel Xeon CPU Max 9480 Processor (1.90 GHz, 56 × 2 cores).

|  | 25 K | 50 K | 75 K | 100 K | 125 K | 150 K | 175 K | 200 K | 250 K | 300 K |
|---|---|---|---|---|---|---|---|---|---|---|
| **QTB-FPMD** | 18.0 | 18.2 | 18.6 | 18.8 | 19.0 | 19.3 | 19.4 | 19.7 | 19.6 | 19.9 |
| **CTB-FPMD** | 17.5 | 19.1 | 19.6 | 20.1 | 20.3 | 22.2 | 21.1 | 21.7 | 22.0 | 22.1 |

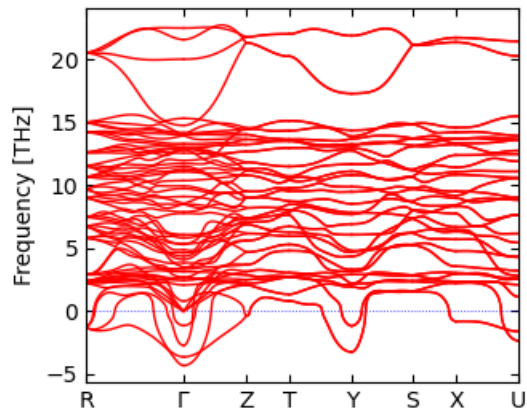

Figure S1. Phonon band structure of the paraelectric phase *Pnma* calculated by PBE_GGA.

Supplemental Materials

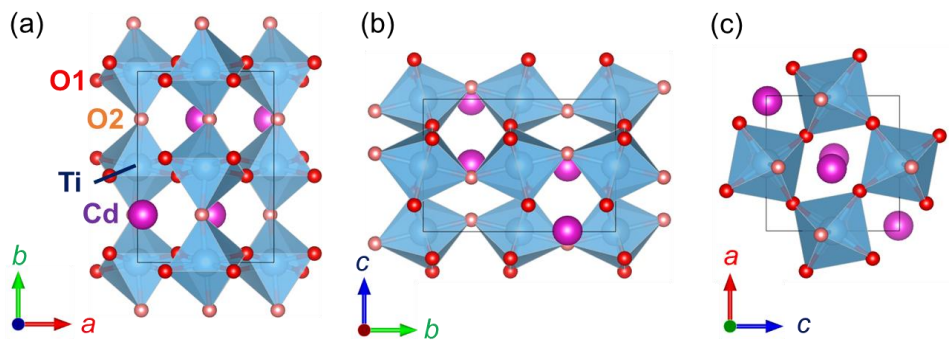

Figure S2. Crystal structure *Pnma* in the (a) *ab*-, (b) *bc*-, and (c) *ca*-plane.

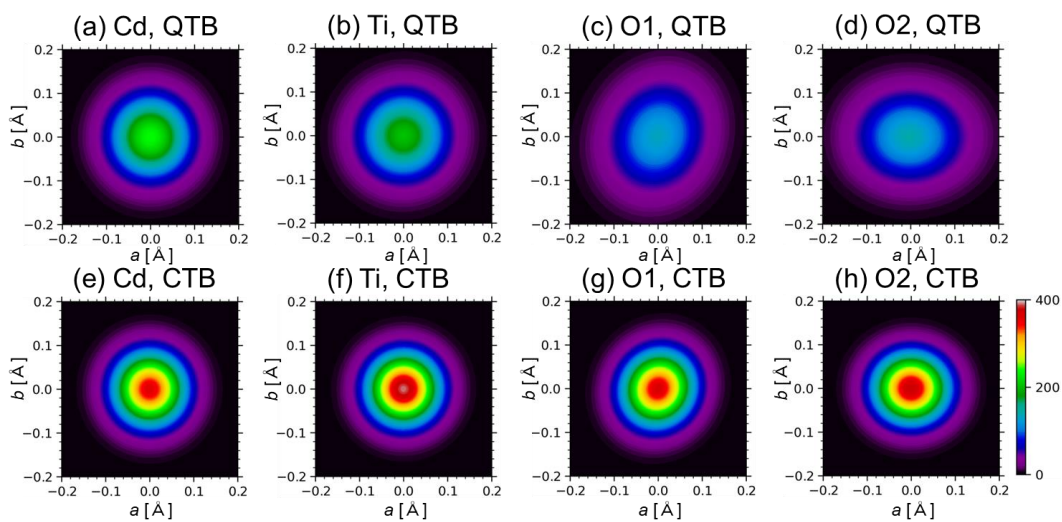

Figure S3. The nuclear density distributions of (a, e) Cd, (b, f) Ti, (c, g) O1 and (d, h) O2 ions in the *ab*-plane during the QTB- and CTB-FPMD simulations at 25 K. The distributions were obtained by accumulating Gaussian distributions centered at each ion with width of 0.05 Å every MD step.

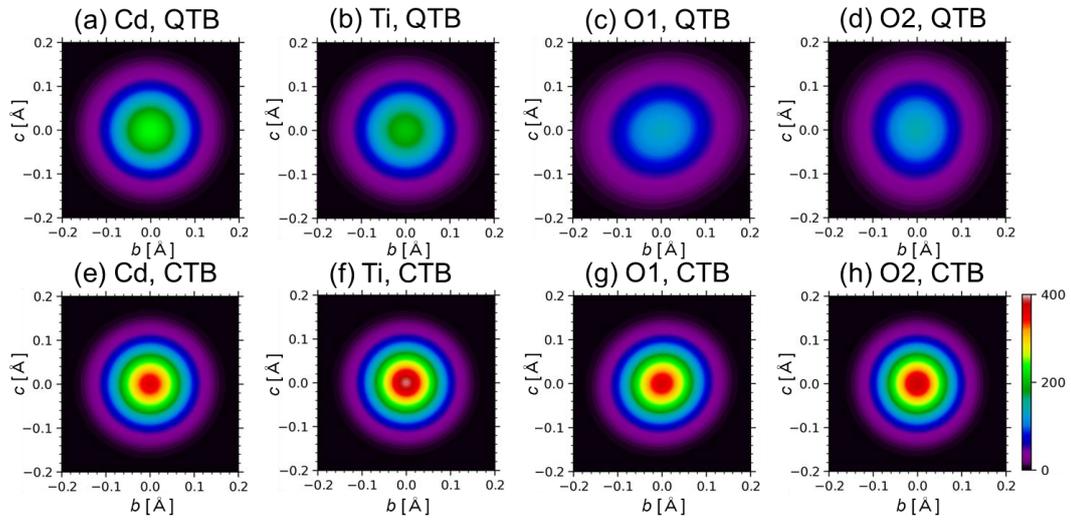

Figure S4. The nuclear density distributions of (a, e) Cd, (b, f) Ti, (c, g) O1 and (d, h) O2 ions in the *bc*-plane during the QTB- and CTB-FPMD simulations at 25 K. The distributions were obtained by accumulating Gaussian distributions centered at each ion with width of 0.05 Å every MD step.

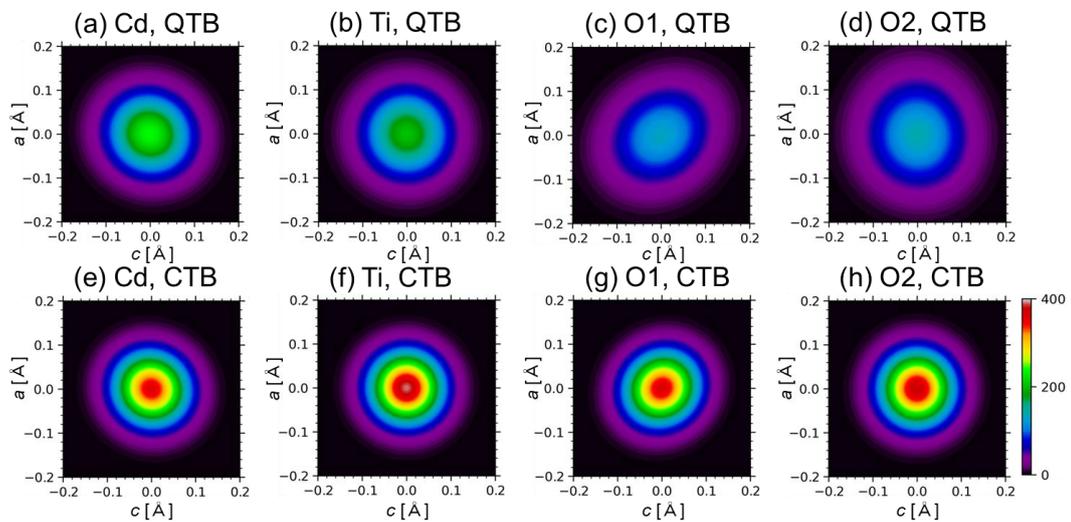

Figure S5. The nuclear density distributions of (a, e) Cd, (b, f) Ti, (c, g) O1 and (d, h) O2 ions in the *ca*-plane during the QTB- and CTB-FPMD simulations at 25 K. The distributions were obtained by accumulating Gaussian distributions centered at each ion with width of 0.05 Å every MD step.